# A two-band model of spin-polarized transport in Fe|Cr|MgO|Fe magnetic tunnel junctions

A. Vedyaev[1,2] , N. Ryzhanova[1,2], N. Strelkov[1,2], M. Chshiev[1] and B. Dieny[1]

[1]*SPINTEC, URA 2512 CEA/CNRS, CEA/Grenoble, 38024 Grenoble Cedex 9, France*

[2]*Department of Physics, Moscow Lomonosov University, Moscow, Russia*

**Abstract**

Theoretical studies of spin dependent transport in Fe/Cr/MgO/Fe tunnel junctions with non-collinear alignment of magnetizations of metallic layers comprising these MTJs are presented. Calculations are performed with use of non-equilibrium Green function technique in the framework of the Keldysh formalism. The Green and wave functions in the barrier region under applied voltage are treated using Wentzel-Kramers-Brillouin (WKB) approximation. Electronic band structure of ferromagnetic electrodes is modeled within a two-band model with majority and minority states being *s*-like and *d*-like electrons, respectively. Furthermore, interfacial *s*-*d* hybridization is taken into account and calculated using perturbation corrections for the wave and Green functions. It is shown that in the presence of Cr layer at the Fe|MgO interface, the contribution from *s*-*d* hybridization to the total current is much stronger in the anti-parallel magnetizations configuration compared to the parallel one leading to decrease of tunnel magnetoresistance (TMR) values in agreement with earlier reports.

## 1. Introduction

Magnetic tunnel junctions (MTJ) have been objects of great interest for both scientists and engineers because of their high sensitivity to magnetic fields, which makes them good candidates for hard drive read heads and magnetic random access memories[1,2] (MRAM). The most attractive among MTJs are their crystalline counterparts due to prediction[3,4] and observation[5,6,7] of large tunnel magnetoresistance (TMR) values due to spin filtering[8] of Bloch states with certain symmetries in Fe|MgO(001) junctions. It was demonstrated using ab-initio calculations[3,4,8,9] that the "*s*-like" band with $\Delta_1$ symmetry (existing at the Fermi energy only for "up" spins in bcc-Fe (Co and CoFe) along (001) direction) has the smallest decay rate MgO due to evanescent state of the same character in this insulator leading to the high conductance in parallel (P) magnetizations configuration of adjacent ferromagnetic electrodes. At the same time, the conductance in anti-parallel (AP) magnetizations configuration yields very small conductance values since there is no "*s*-like" band for "down" spins in aforementioned bcc ferromagnets but "*d*-like" bands with $\Delta_2$, $\Delta_{2'}$ and $\Delta_5$ symmetries.

Recent exciting experiments[10] demonstrated that additional Cr layer of thickness *a* inserted at the interface between Fe and MgO layers can be used as a tunnel "barrier" for *s*-like electronic states with $\Delta_1$ symmetry providing straightforward evidence of Bloch states symmetry –based spin filtering. In certain sense, bcc-Fe (Co and CoFe) in MgO-based crystalline MTJs should demonstrate "half-metallic"- like behavior leading to huge TMR values. However, in aforementioned experiment[10] with Fe|Cr|MgO|Fe the TMR strongly decreased as a function of Cr thickness and the highest experimental value reported so far in CoFe|MgO|CoFe structures is 600% at room temperature[11]. One of the possible reasons preventing infinite TMR values can be appearance of the partial layer of FeO at the interface which represents an additional barrier for *s*-electron[8,12]. Another reason could be a symmetry break at the Fe|MgO and/or Cr|MgO interface which in turn leads to the *s*-*d* hybridization with transition to the conducting state in the anti-parallel configuration of magnetizations in Fe electrodes.

In this work we will theoretically investigate the influence of *s*-*d* hybridization on TMR in Fe/Cr/MgO/Fe and Fe/MgO/Fe structures with noncollinear alignment of magnetizations of ferromagnetic layers comprising these MTJs.

## 2. Model

For sake of simplicity and in a view of arguments stated above, we will describe electronic structure of ferromagnetic electrodes within two bands model, i.e. with majority *s*-like electrons and minority *d*-like holes. It follows from this model that *s*-electron as well as *d*-hole bands have zero conductivity in the case of the anti-parallel configuration of magnetizations in the outer ferromagnetic layers, so the TMR in this structure has to reach infinite value. To calculate the conductivity in this non-collinear structure we will use the Keldysh technique for non-equilibrium Green functions[13] with the perturbation corrections coming from *s*-*d* hybridization. WKB approximation will be used for wave and Green functions in the trapezoidal barrier region under applied voltage[14]. To improve further the model we consider different effective mass for electron and hole bands in each layer. The details of this approach modified for tunnel transport could be found in Ref. [14]. To take into account *s*-*d* hybridization one need to find corresponding corrections to the wave functions using the following matrix equation:

$$\begin{pmatrix}\psi_s^{\uparrow}(z)\\ \psi_s^{\downarrow}(z)\\ \psi_d^{\uparrow}(z)\\ \psi_d^{\downarrow}(z)\end{pmatrix} = \begin{pmatrix}\psi_{0s}^{\uparrow}(z)\\ \psi_{0s}^{\downarrow}(z)\\ \psi_{0d}^{\uparrow}(z)\\ \psi_{0d}^{\downarrow}(z)\end{pmatrix}$$
$$+\begin{pmatrix} 0 & 0 & \sum_i G_s^{\uparrow\uparrow}(z,z_i)\gamma_i^{\uparrow\uparrow} & \sum_i G_s^{\uparrow\downarrow}(z,z_i)\gamma_i^{\downarrow\downarrow}\\ 0 & 0 & \sum_i G_s^{\downarrow\uparrow}(z,z_i)\gamma_i^{\uparrow\uparrow} & \sum_i G_s^{\downarrow\downarrow}(z,z_i)\gamma_i^{\downarrow\downarrow}\\ \sum_i G_d^{\uparrow\uparrow}(z,z_i)\gamma_i^{\uparrow\uparrow} & \sum_i G_d^{\uparrow\downarrow}(z,z_i)\gamma_i^{\downarrow\downarrow} & 0 & 0\\ \sum_i G_d^{\downarrow\uparrow}(z,z_i)\gamma_i^{\uparrow\uparrow} & \sum_i G_d^{\downarrow\downarrow}(z,z_i)\gamma_i^{\downarrow\downarrow} & 0 & 0 \end{pmatrix} \times \begin{pmatrix}\psi_s^{\uparrow}(z)\\ \psi_s^{\downarrow}(z)\\ \psi_d^{\uparrow}(z)\\ \psi_d^{\downarrow}(z)\end{pmatrix}, \quad (1)$$

where $\psi_{0s(d)}^{\uparrow(\downarrow)}(z)$, $\psi_{s(d)}^{\uparrow(\downarrow)}(z)$ are zero order on hybridization and full wave functions for $s$, $d$ carriers and for "up", "down" spins, $G_{s(d)}^{\sigma\sigma'}(z,z_i)$ – retarded Green functions for $s$, $d$ carriers, $z_i$ – is the metal/barrier interfacial coordinate (the $z$ axis is perpendicular to the interfaces), and $\gamma_i^{\sigma\sigma}$ – are the parameters of hybridization on the $i$-th interface. It follows from (1) that the most important is assumed to be the hybridization at the metal/insulator interface. In general, there are also terms proportional to the off-diagonal on spin parameters of hybridization due to the spin-orbit coupling which is neglected here so it's reasonable to assume that $\gamma_i^{\sigma,-\sigma} \ll \gamma_i^{\sigma\sigma}$. All wave and Green functions depend on energy $E$ of tunneling carriers and on in-plane momentum $\vec{\kappa}$ but for simplicity we will omit these dependences in expressions below. To find the current we have to build the non-equilibrium Green function:

$$G_{\sigma\sigma\kappa}^{-+}(z,z') = \mathrm{f}(E)\psi_l^{\sigma\sigma\,*}(z')\psi_l^{\sigma\sigma}(z) + \mathrm{f}(E-eV)\psi_r^{\sigma\sigma\,*}(z')\psi_r^{\sigma\sigma}(z), \tag{2}$$

where indexes $l$, $r$ mean that that they are the functions for left-right and right-left moving carriers correspondently, $\mathrm{f}(E)$, $\mathrm{f}(E-eV)$ – are the Fermi distributions in left, right electrodes, and $\vec{\kappa}$ is the momentum along the interface. Then the current is calculated using the formula:

$$j^{\sigma} \propto i \iint \kappa d\kappa dE \left(\frac{\partial}{\partial z'} - \frac{\partial}{\partial z}\right) G_{\sigma\sigma\kappa}^{-+}(z,z')\Big|_{z=z'} \tag{3}$$

The zero order spin "up" $s$-current and spin "down" d-current are given by the integrals:

$$j_{0s}^{\uparrow} = C \iint \kappa d\kappa dE\, j_{0s\kappa}^{\uparrow}(E); \quad j_{0d}^{\downarrow} = C \iint \kappa d\kappa dE\, j_{0d\kappa}^{\downarrow}(E), \tag{4}$$

with corresponding integrands defined as:

$$\begin{aligned} j_{0s\kappa}^{\uparrow}(E_e^s) = \frac{q_1 q_2 \operatorname{Re} k_1 \, |k_3|^2 m_1 m_3^2 m_b^2}{|den|^2} \Big[ &(1 + \cos\theta) \operatorname{Re} k_4 \, m_1 \big| K(q_2 m_2 - i k_5 m_b)\varphi_2^{\downarrow} \\ &- K^{-1}(q_2 m_2 + i k_5 m_b)\varphi_4^{\downarrow}\big|^2 + (1 - \cos\theta) \operatorname{Re} k_5 \, m_2 \big| K(q_2 m_1 - i k_4 m_b)\varphi_2^{\downarrow} \\ &- K^{-1}(q_2 m_1 + i k_4 m_b)\varphi_4^{\downarrow}\big|^2 \Big], \end{aligned} \tag{5}$$

$$j^{\downarrow}_{0d\kappa}\left(E^d_e\right) = \frac{q^d_1 q^d_2 \,\mathrm{Re}\, k^d_2 \left|k^d_3\right|^2 m^d_2 m^{d\,2}_3 m^{d\,2}_b}{|den|^2_d}\Big[(1 - \cos\theta)\,\mathrm{Re}\, k^d_4\, m^d_1 \big|K(q_2 m_2 - ik_5 m_b)\varphi^{\uparrow}_2 - K^{-1}(q_2 m_2 + ik_5 m_b)\varphi^{\uparrow}_4\big|^2_d + (1 + \cos\theta)\,\mathrm{Re}\, k^d_5\, m^d_2 \big|K(q_2 m_1 - ik_4 m_b)\varphi^{\uparrow}_2 - K^{-1}(q_2 m_1 + ik_4 m_b)\varphi^{\uparrow}_4\big|^2_d\Big]$$

where

$$K = \exp\int_{z_2}^{z_3} q(z)\,dz = \exp\left[\frac{b}{3m_b \times 0.288V}\left\{(q_0^2 + \kappa^2 + 0.288\varepsilon)^{3/2} - \left(q_0^2 + \kappa^2 - 0.288(V - \varepsilon)\right)^{3/2}\right\}\right];\; q^2(z) = \frac{2m_b}{\hbar^2}\left(U - E_e - eV\frac{z - z_2}{z_3 - z_2}\right) + \kappa^2;\; q_{1(2)} = q\left(z_{2(3)}\right)$$

In expressions above $z_2$, $z_3$ represent coordinates of the Cr/MgO and MgO/Fe interfaces, respectively; $V$ is voltage, $\theta$ is the angle between magnetizations of Fe layers, $C = 0.1385 m_1^{-1} \times 10^{13}$ A/cm$^2$, $m_1$, $m_2$ – are normalized effective mass for *s*-electron with “up”, “down” spin in Fe electrodes, $m_3$, $m_b$ are the same in Cr and the barrier layers, and 0.288 is factor converging the units of energies from $eV$ into Å$^{-2}$. Furthermore,

$$den = (1 + \cos\theta)\left[E(q_2 m_1 - ik_4 m_b)\varphi^{\uparrow}_2 - E^{-1}(q_2 m_1 + ik_4 m_b)\varphi^{\uparrow}_4\right]\left[E(q_2 m_2 - ik_5 m_b)\varphi^{\downarrow}_2 - E^{-1}(q_2 m_2 + ik_5 m_b)\varphi^{\downarrow}_4\right] + (1 - \cos\theta)\left[E(q_2 m_2 - ik_5 m_b)\varphi^{\uparrow}_2 - E^{-1}(q_2 m_1 + ik_5 m_b)\varphi^{\uparrow}_4\right]\left[E(q_2 m_1 - ik_4 m_b)\varphi^{\downarrow}_2 - E^{-1}(q_2 m_1 + ik_4 m_b)\varphi^{\downarrow}_4\right]$$

$$\varphi_2^{\uparrow(\downarrow)} = e^{ik_3 a}\left(m_{1(2)} k_3 - k_{1(2)} m_3\right)(q_1 m_3 + ik_3 m_b) + e^{-ik_3 a}\left(m_{1(2)} k_3 + k_{1(2)} m_3\right)(q_1 m_3 - ik_3 m_b);$$

$$\varphi_4^{\uparrow(\downarrow)} = e^{ik_3 a}\left(m_{1(2)} k_3 - k_{1(2)} m_3\right)(q_1 m_3 - ik_3 m_b) + e^{-ik_3 a}\left(m_{1(2)} k_3 + k_{1(2)} m_3\right)(q_1 m_3 + ik_3 m_b);$$

where $k_1$, $k_4$ ($k_2$, $k_5$) indicate *z*-components of *s*-electron momentum with energy $E$ for “up”, “down” spins in the left (right) electrodes, and $k_3$ is the same Cr layer:

$$k^2_{1(2)} = \frac{2m_{1(2)}}{\hbar^2}(E_e - U_s \pm J_{sd}) - \kappa^2;$$

$$k^2_{4(5)} = \frac{2m_{1(2)}}{\hbar^2}(E_e - U_s \pm J_{sd} + eV) - \kappa^2;$$

$$k^2_3 = \frac{2m_3}{\hbar^2}(E_e - U_s^{Cr}) - \kappa^2.$$

Here $U_s$ – is the bottom of *s*-band in Fe without exchange splitting $J_{sd}$, $U_s^{Cr}$ - the same for Cr layer.

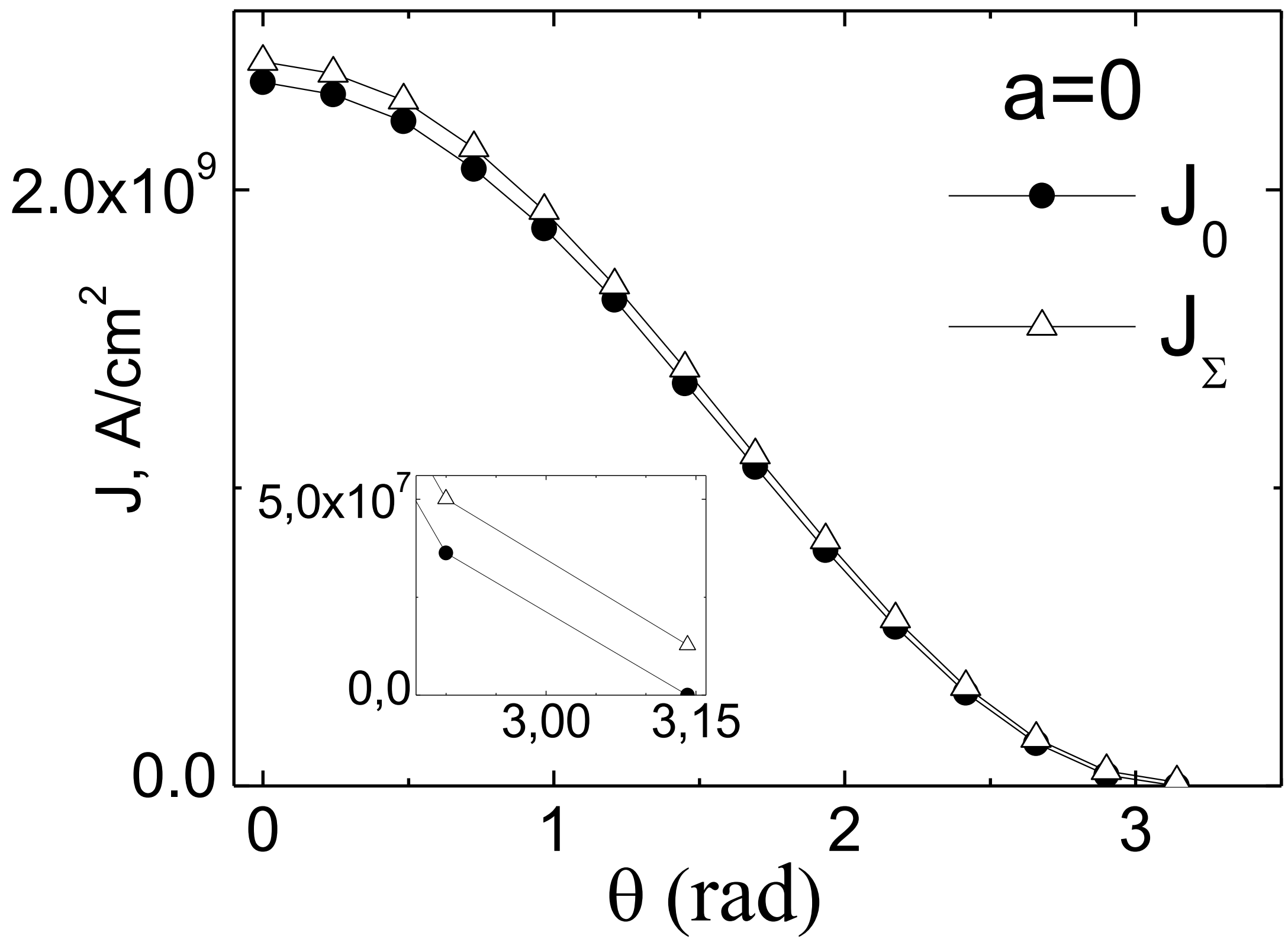

**Fig. 1.** Angular dependence of the current in Fe|MgO|Fe tunnel junction with no ($j_0$) and with *s*-*d* hybridization ($j_\Sigma$) taken into account. Inset shows a zoom around θ=π.

Note that in our model the bottom of electron *s*-band is situated above the Fermi level giving rise to purely imaginary value for $k_3$. As it follows from **Error! Reference source not found.** and **Error! Reference source not found.** for purely imaginary $k_5$ and $k_4^d$ (which is the case in our model) and $\theta = \pi$, both *s*- and *d*-currents vanishes giving $TMR \to \infty$.

In the presence of *s*-*d* hybridization additional terms in wave functions appear, for example:

$$\Delta\psi_s^{\uparrow(\uparrow)}(z < z_1) = G_s^{\uparrow\uparrow}(z, z_1)\gamma_i^{\uparrow\uparrow} G_d^{\uparrow\uparrow}(z_i, z_i)\gamma_i^{\uparrow\uparrow}\psi_s^{\uparrow\uparrow}(z_i)$$

Taking into account additional terms of the second order on *s*-*d* hybridization, one gets four new currents: two additional ones for *s*-electron currents due to *s*-*d* hybridization at Cr|MgO and MgO|Fe interfaces and similarly the two for *d*-holes: $\Delta j_2^{s\uparrow} + \Delta\tilde{j}_2^{s\uparrow}$, $\Delta j_3^{s\uparrow} + \Delta\tilde{j}_3^{s\uparrow}$, $\Delta j_2^{d\downarrow} + \Delta\tilde{j}_2^{d\downarrow}$ and $\Delta j_3^{d\downarrow} + \Delta\tilde{j}_3^{d\downarrow}$. For example, the corrections from the Cr|MgO interface are written as $\Delta j_{2\kappa}^{s\uparrow} +$

$\Delta \tilde{j}_{2\kappa}^{s\uparrow}$ given by the following expressions (we write down here the integrands, and the limits of integration were chosen from the condition of reality *z*-momentum for *s*- and *d*-carriers respectively):

$$\Delta j_{2\kappa}^{s\uparrow} = 16 m_1 m_3^2 m_b^2 m_3^d m_b^d k_1 \left(\gamma_2^{\uparrow\uparrow}\right)^2 \operatorname{Im}\left[r_1^{\uparrow *} k_3^2 \left(\phi_{2s}^{\uparrow(\uparrow)}\right)^2 \phi_{2d}^{\uparrow(\uparrow)} \left\{e^{ik_3^d a}\left(m_1^d k_3^d - k_1^d m_3^d\right) + e^{-ik_3^d a}\left(m_1^d k_3^d + k_1^d m_3^d\right)\right\}\right]; r_1^{\uparrow} = \frac{num^{\uparrow}}{den}; \tag{6}$$

where

$$\begin{aligned} num^{\uparrow} = {} & (1+\cos\theta)\left[E(q_2 m_1 - ik_4 m_b)\varphi_1^{\uparrow} - E^{-1}(q_2 m_1 + ik_4 m_b)\varphi_3^{\uparrow}\right] \\ & \times \left[E(q_2 m_2 - ik_5 m_b)\varphi_2^{\downarrow} - E^{-1}(q_2 m_2 + ik_5 m_b)\varphi_4^{\downarrow}\right] \\ & + (1-\cos\theta)\left[E(q_2 m_2 - ik_5 m_b)\varphi_1^{\uparrow} - E^{-1}(q_2 m_2 + ik_5 m_b)\varphi_3^{\uparrow}\right] \\ & \times \left[E(q_2 m_1 - ik_4 m_b)\varphi_2^{\downarrow} - E^{-1}(q_2 m_1 + ik_4 m_b)\varphi_4^{\downarrow}\right]; \end{aligned}$$

$$\begin{aligned} \varphi_1^{\uparrow,\downarrow} = {} & e^{ik_3 a}\left(m_{1,2} k_3 + k_{1,2} m_3\right)(q_1 m_3 + ik_3 m_b) \\ & + e^{-ik_3 a}\left(m_{1,2} k_3 - k_{1,2} m_3\right)(q_1 m_3 - ik_3 m_b); \end{aligned}$$

$$\begin{aligned} \varphi_3^{\uparrow,\downarrow} = {} & e^{ik_3 a}\left(m_{1,2} k_3 + k_{1,2} m_3\right)(q_1 m_3 - ik_3 m_b) \\ & + e^{-ik_3 a}\left(m_{1,2} k_3 - k_{1,2} m_3\right)(q_1 m_3 + ik_3 m_b) \end{aligned}$$

$$\begin{aligned} \phi_{2s}^{\uparrow(\uparrow)} = {} & \frac{1}{den}\left\{(1+\cos\theta)\left[E(q_2 m_2 - ik_5 m_b)\varphi_2^{\downarrow} - E^{-1}(q_2 m_2 + ik_5 m_b)\varphi_4^{\downarrow}\right]\right. \\ & \times \left[E(q_2 m_1 - ik_4 m_b) + E^{-1}(q_2 m_1 + ik_4 m_b)\right] \\ & + (1-\cos\theta)\left[E(q_2 m_1 - ik_4 m_b)\varphi_2^{\downarrow} - E^{-1}(q_2 m_1 + ik_4 m_b)\varphi_4^{\downarrow}\right] \\ & \left.\times \left[E(q_2 m_2 - ik_5 m_b) + E^{-1}(q_2 m_2 + ik_5 m_b)\right]\right\} \end{aligned}$$

$$\begin{aligned} \Delta \tilde{j}_{2\kappa}^{\uparrow}(E_s) = {} & -256 q_1 q_2 q_1^d q_2^d m_b^3 m_3^3 \left(m_b^d m_3^d\right)^2 m_1 k_1 \sin^2\theta\, \gamma_2^{\uparrow\uparrow}\gamma_2^{\downarrow\downarrow} \\ & \times \operatorname{Im}\left\{\frac{r_1^{\uparrow *} k_3^2 \phi_2^{\uparrow(\uparrow)}}{den \times den^d}(k_4 m_2 - k_5 m_1)\left(k_4^d m_2^d - k_5^d m_1^d\right)\left[e^{ik_3 a}(k_3 m_2 - k_2 m_3)\right.\right. \\ & \left. + e^{-ik_3 a}(k_3 m_2 + k_2 m_3)\right] \\ & \times \left[e^{ik_3^d a}\left(k_3^d m_2^d - k_2^d m_3^d\right) + e^{-ik_3^d a}\left(k_3^d m_2^d + k_2^d m_3^d\right)\right] \\ & \left.\times \left[e^{ik_3^d a}\left(k_3^d m_1^d - k_1^d m_3^d\right) + e^{-ik_3^d a}\left(k_3^d m_1^d + k_1^d m_3^d\right)\right]\right\} \end{aligned}$$

Finally, the total current is given by the sum of all calculated zero order and additional integrated currents:

$$J_\Sigma = j_{0s}^{\uparrow} + j_{0d}^{\downarrow} + \Delta j_2^{s\uparrow} + \Delta \tilde{j}_2^{s\uparrow} + \Delta j_3^{s\uparrow} + \Delta \tilde{j}_3^{s\uparrow} + \Delta j_2^{d\downarrow} + \Delta \tilde{j}_2^{d\downarrow} + \Delta j_3^{d\downarrow} + \Delta \tilde{j}_3^{d\downarrow} \quad (7)$$

## 5. Results and discussion

In order to demonstrate the impact of interfacial *s-d* hybridization using analytical results of the previous section, we calculated the angular dependence of both zero-order and additional currents using Eqs (4) and (7), respectively. The results of this calculation for Fe|MgO|Fe and Fe|Cr|MgO|Fe structures are shown in Figs. 1 and 2. In Fig. 1 we show the dependence of the tunnel current across the standard Fe|MgO|Fe (Cr thickness a=0) junction as a function of angle between two Fe magnetizations. As one can see the tunnel current $J_0$ (filled circles) in the absence of *s-d* hybridization has usual angular dependence and vanishes for antiparallel (AP) alignment of magnetizations (θ=π) according to the half-metallic like picture according to our model with infinite TMR. At the same time, the total current $J_\Sigma$ (open triangles) which takes into account the second order contributions due to *s-d* hybridization is not zero for AP configuration (see inset in Fig. 1). This happens because propagating majority (minority) *s* (*d*) electrons are converted at the interface into *d* (*s*) electrons and continues to propagate freely in the layer with opposite direction of the magnetization. The same dependences for the junction with finite Cr layer thickness are presented in Fig. 2. We notice a strong decrease in the parallel current value since the Cr represents an additional barrier for *s*-electrons (which is part of $\Delta_1$ Bloch state). In addition, the role of *s-d* hybridization is further enhanced leading to significant drop of TMR ratio.

We chose the model simulating the half metal nature for both *s* and *d*-electrons in the absence of hybridization to show more clearly the crucial influence of this hybridization on TMR. We have demonstrated that due to the *s-d* mixing the additional currents comparing with the incident *s*-up current appear. Due to these processes inside the barrier $j_d^{\uparrow}$ current even dominate the $j_s^{\uparrow}$ one because the $\psi_s^{\uparrow}$ states easily penetrate it. For the same reason in the region 3 in the vicinity of the interface the current $j_d^{\downarrow}$ almost equal to the current because *s*-electron on the interface transform in the *d* one, which according to the chosen half-metallic model easily propagates in the ferromagnetic.

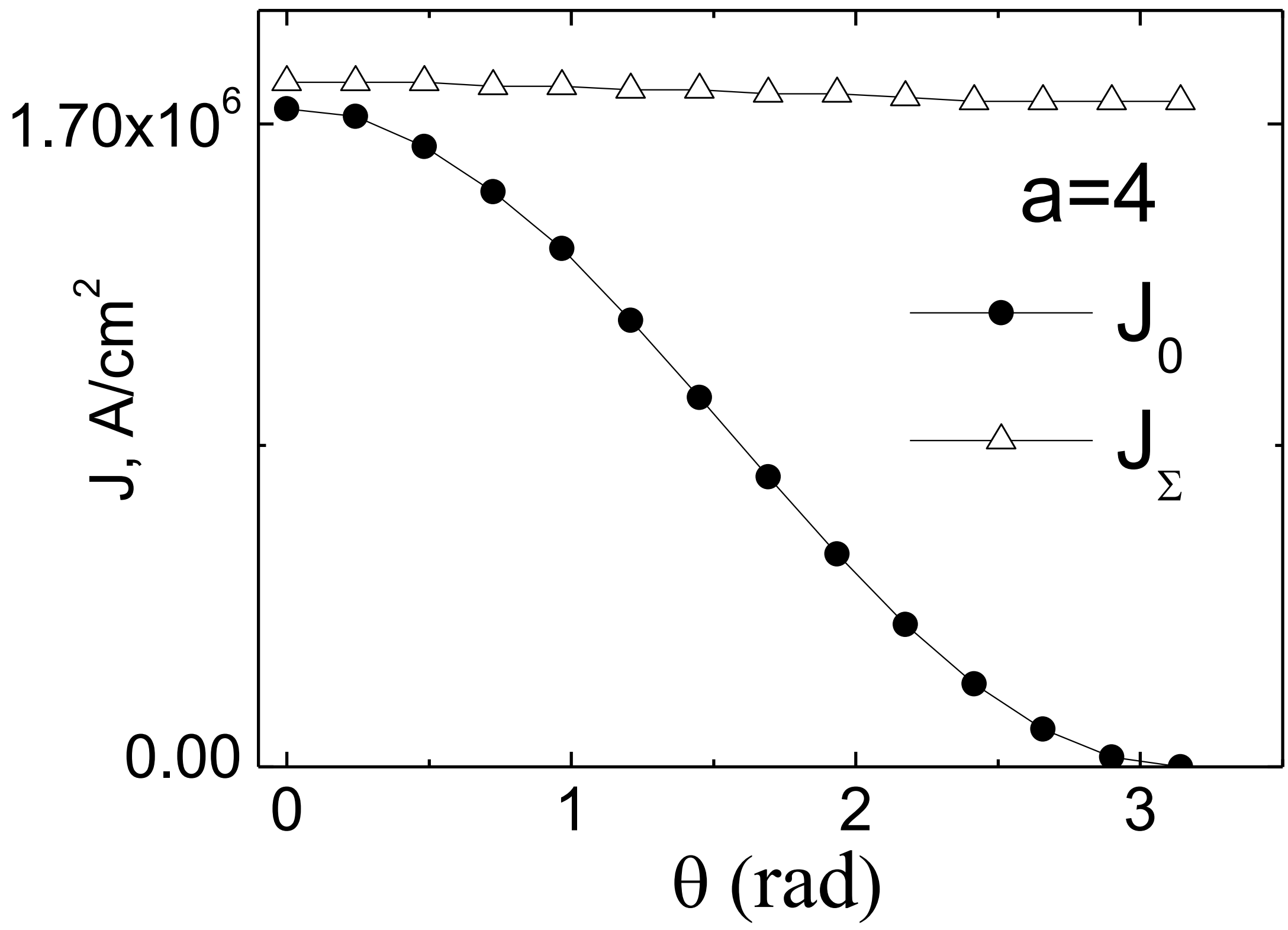

**Fig. 2.** The same dependencies as in Fig. 1 for Fe|Cr|MgO|Fe tunnel junction with finite Cr layer thickness

This work was partly supported by RFFI grant No. 07-02-00918 (Russia) and by Nanosciences Foundation Chair of Excellence Program (Grenoble, France).

[1] S. S. P. Parkin, K. P. Roche, M. G. Samant et al., J. of Appl. Phys., 85, 5828 (1999)

[2] S. Tehrani, E. Chen, M. Durlam et al, J. of Appl. Phys., 85, 5822 (1999)

[3] W. H. Butler et al, Phys. Rev. B, 63, 054416 (2001)

[4] J. Mathon and A. Umerski, Phys. Rev. B, 63, 220403(R) (2001)

[5] S. S. P. Parkin et al, Nature Materials 3, 862 (2004)

[6] S. Yuasa et al, Nature Materials 3, 868 (2004)

[7] J. Faure-Vincent et al, Appl. Phys. Lett. 82, 4507 (2003)

[8] W. H. Butler et al, IEEE Trans. Magn. 41, 2645 (2005)

[9] X.-G. Zhang and W. H. Bulter, Phys. Rev. B. 70, 172407 (2004)

[10] F. Greullet, C. Tiusan, F. Montaigne, M. Hehn, D. Halley, O. Bengone, M. Bowen and W. Weber, Phys. Rev. Lett. 99,187202 (2007)
[11] S. Ikeda et al., Appl. Phys. Lett. 93, 082508 (2008)
[12] S. Yuasa, A. Fukushima,T. Nagahama, K. Ando, Y. Suzuki, J. Appl. Phys. 43, L588 (2004)
[13] L. Keldysh, Soviet Phys. JETP, 20, 1018( 1965)
[14] A. Manchon, N. Ryzhanova, A. Vedyaev, M. Chshiev and B. Dieny, J. Phys.: Cond. Mat. 20, 145208 (2008)